# An Open-Source Data Storage and Visualization Platform for Collaborative Qubit Control

Devanshu Brahmbhatt[1,2,3], Yilun Xu[1,*], Neel Vora[1,2,3], Larry Chen[4], Neelay Fruitwala[1], Gang Huang[1], Qing Ji[1], and Phuc Nguyen[2,*]

[1]Lawrence Berkeley National Laboratory, Berkeley, CA, 94720, USA
[2]University of Massachusetts Amherst, College of Information and Computer Sciences, Amherst, MA, 01003, USA
[3]The University of Texas at Arlington, Computer Science and Engineering, Arlington, TX, 76010, USA
[4]University of California, Berkeley, Department of Physics, Berkeley, CA, 94720, USA
*Co-corresponding authors: Yilun Xu (email: yilunxu@lbl.gov) and Phuc Nguyen (email: vp.nguyen@cs.umass.edu)

## ABSTRACT

Developing collaborative research platforms for quantum bit control is crucial for driving innovation in the field, as they enable the exchange of ideas, data, and implementation to achieve more impactful outcomes. Furthermore, considering the high costs associated with quantum experimental setups, collaborative environments are vital for maximizing resource utilization efficiently. However, the lack of dedicated data management platforms presents a significant obstacle to progress, highlighting the necessity for essential assistive tools tailored for this purpose. Current qubit control systems are unable to handle complicated management of extensive calibration data and do not support effectively visualizing intricate quantum experiment outcomes. In this paper, we introduce *QubiCSV* (Qubit Control Storage and Visualization), a platform specifically designed to meet the demands of quantum computing research, focusing on the storage and analysis of calibration and characterization data in qubit control systems. As an open-source tool, *QubiCSV* facilitates efficient data management of quantum computing, providing data versioning capabilities for data storage and allowing researchers and programmers to interact with qubits in real time. The insightful visualization are developed to interpret complex quantum experiments and optimize qubit performance. *QubiCSV* not only streamlines the handling of qubit control system data but also improves the user experience with intuitive visualization features, making it a valuable asset for researchers in the quantum computing domain.

## 1 Introduction

Quantum computing, a significant leap from classical computing, utilizes the principles of quantum mechanics to potentially solve problems that are challenging for traditional computers[1]. At its core, the quantum bit or qubit[2], serves as the fundamental unit of quantum information, differing from classical bits[3] in its ability to exist in multiple states simultaneously due to quantum superposition[4]. This unique property enables quantum computers to process information in ways that classical systems cannot, making them particularly effective for solving certain complex problems[5]. In order to unlock the complete capabilities of quantum computing, mastering a pivotal concept known as qubit control[6] is imperative. Qubit control encompasses the meticulous manipulation of qubits for executing quantum operations, a task that demands exceptional precision due to qubits' susceptibility to noise[7,8] and decoherence[9].

In current quantum computing, especially superconducting qubit research, readout plays a fundamental role by translating quantum information into classical information, which is then represented within the computational framework as binary digits. Qubit readout stands out as one of the most challenging and time-consuming tasks on superconducting quantum processors due to its susceptibility to errors and slow execution. Field-Programmable Gate Arrays (FPGAs) offer a promising solution for qubit readout due to their adaptability and ability to process multiple signals simultaneously in parallel. This capability enables FPGAs to manage the complexities of qubit readout effectively, reducing errors and enhancing efficiency. The FPGA-based solution has been used to generate radio frequency (RF) pulses that alter the qubit state in qubit control systems[7], and subsequently measured for output [10]. State-of-the-art qubit control systems include Zurich Instruments[11], Quantum Machines[12], Keysight[13], Qblox[14], and QICK[15]. Among these, the QubiC (Qubit Control)[16,17] system, developed at Lawrence Berkeley National Laboratory (LBNL), represents a notable open-source advancement enabling mid-circuit measurement and feed-forward[18]. QubiC's open-source nature, high performance, and modular design align it closely with the needs of quantum research, positioning it as a valuable tool for scientists in navigating the evolving quantum computing domain.

Figure 1 depicts an actual configuration of a quantum computer at the Quantum Nanoelectronics Laboratory[19], University of California, Berkeley, which includes a dilution fridge, a critical component in quantum computing where superconducting qubits are housed and operated at extremely low temperatures. Accompanying the fridge are multiple qubit control systems, each meticulously crafted to manage and manipulate the qubits.

Quantum information processors require frequent calibration and characterization to mitigate drifts and errors[20,21] arising from environmental fluctuations and hardware imperfections, thus ensuring reliable and accurate computation[22]. Calibration[23] and characterization data storage are critical components in quantum experimentation. Hence, characterization data storage provides a repository for experimental results, enabling scientists to track the performance and stability of quantum systems over time. This data are indispensable for ongoing research, allowing for the analysis of trends, the identification of anomalies, and the refinement of quantum models and algorithms.

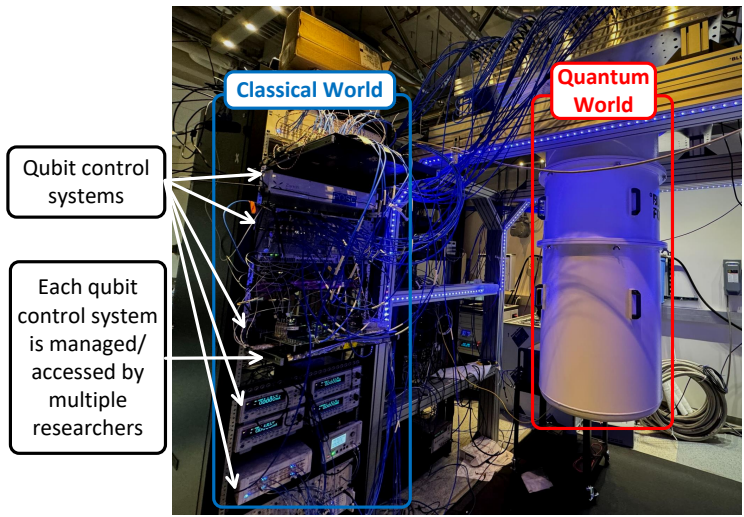

**Figure 1.** An example setting of quantum control system research

Maintaining and updating these calibration files, which encompass settings for numerous gates and qubits, poses a formidable challenge for scientists. Each team member (researchers, engineers, students), focusing on different aspects of the quantum system, generates their unique calibration files, which need to be accurately tracked and updated. This task becomes increasingly challenging as the frequency of calibration increases. Furthermore, the need for collaboration among scientists adds another layer of complexity. Sharing these extensive files and ensuring that all team members have access to the latest versions is a logistical hurdle, slowing efficient collaboration and progress as sharing and syncing these extensive files without a reliable versioning system can restrict efficient teamwork and slow down progress. Furthermore, during post-experimentation, the qubit control system generates an experimental result file, for example, *chip_name.data.json* in QubiC[16,17], which holds big potential for providing insights into the experiment's outcomes. However, the absence of a dedicated storage solution and a method to save this file directly from the hardware limits its utility. Scientists are left with a wealth of data but lack the means to store, manage, and analyze it effectively.

Visualization plays an important role in enhancing the understanding and monitoring of quantum experiments. It allows researchers to observe and analyze complex Quantum calibration and characterization data in an intuitive manner. Through visualization, patterns, and insights that might otherwise remain obscured in raw data can be brought to light, facilitating a deeper understanding of quantum phenomena. Initiatives such as VACSEN[24] and QVis[25] focused on the visualization of quantum errors and noise. These tools have laid a foundational framework for understanding quantum system behaviors, which our platform builds upon and expands. Additionally, IBM 's[26] work with the IBMQ calibration database represents another pivotal contribution, offering a comprehensive approach to managing and visualizing calibration data in quantum computing systems. These developments[27,28] collectively inform and inspire our approach. However, they only support noise and error visualization, which, while crucial, does not encompass the entire scope of data visualization needs in quantum computing research.

*QubiCSV*. To overcome the aforementioned limitations, we designed QubiCSV, an end-to-end platform for real-time data management and visualization for quantum computing. We tailor our parameters to align with the QubiC system[16,17], although configurations can be adjusted at the software level to ensure compatibility with alternative qubit control systems. QubiCSV's unique approach to data storage - utilizing data versioning similar to Git [29] but implemented in a database—provides a novel way to manage Quantum calibration and characterization data. This method allows for more effective tracking and reverting of changes over time, which is especially beneficial in a field where experiments are frequently adjusted and iterated upon. As a holistic platform, QubiCSV enhances data visualization in quantum computing research, providing dynamic and interactive tools for visualizing calibration and experiment outcome data. These features aid not only in identifying optimal calibration but also improve the understanding of qubit performance, thereby boosting productivity. In terms of performance, QubiCSV's design is both scalable and user-friendly. Its adaptability and open-source nature facilitate easy integration with various quantum control systems, making it an efficient solution for the evolving needs of research.

In summary, this paper makes the following contributions. We identify data management and visualization challenges in classical-quantum computer systems. Addressing these challenges enhances collaboration and productivity in quantum research. We introduce real-time data management techniques, incorporating data versioning for efficient interaction with qubit systems. This feature streamlines research processes and enables dynamic data management. We present a visualization technique for on-the-fly analysis of calibration and experiment results, facilitating the optimization of calibration configurations and enhancement of qubit performance. We assess the platform at LBNL with a diverse team of researchers, demonstrating its benefits and identifying limitations. This offers valuable insights for future improvements and broader applications.

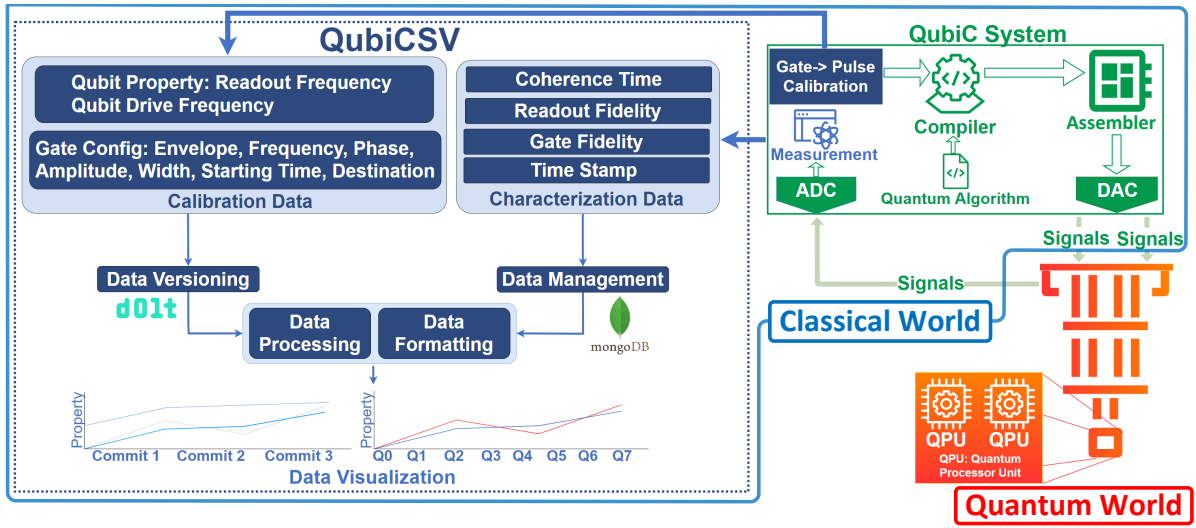

**Figure 2.** QubiCSV's concept: an overview of the data management and visualization for qubit control system research. In the current superconducting qubit control design where the quantum algorithm is transformed into a pulse-level program within the compiler, integrating the calibration configuration. Subsequently, the assembler converts it into binaries, which are then loaded onto an FPGA. From there, the data is translated into RF signals through a digital-to-analog converter (DAC) before being transmitted to a dilution refrigerator housing chips (quantum processors) embedded with superconducting qubits. Once the experiment is completed, the results are captured by an analog-to-digital converter (ADC) and post-processed, then stored in a MongoDB database for visualization. If the calibration proves satisfactory for the experiments, researchers can then store the calibration file in their desired branch on the Dolt database, making it accessible for further visualization and analysis.

## 2 Results

We implemented QubiCSV, which represents one of the first efforts in Quantum calibration and characterization data management and visualization. By implementing a versioning database tailored for qubit control devices, QubiCSV addresses the pressing need for improved collaboration and effective data management. This feature is particularly crucial, given the complexity and dynamism of Quantum calibration and characterization data. Moreover, QubiCSV's real-time data visualization capabilities significantly enhance researchers' ability to interpret complex quantum experiments and optimize qubit performance as illustrated in Figure 2. QubiCSV enables seamless interaction between users to control the system in the physical world to manipulate/control the superconducting qubits in the quantum world.

### 2.1 QubiCSV Data Management

QubiCSV design is inspired by Model-View-Controller (MVC) architecture[30,31], with the model managing database queries and returning data as requested by the controller[32]. The view then renders this data, presenting it to the users in an accessible format. This system is built on the team members' daily routines, with a focus on their utilization of the Jupyter Notebook for organizing code and data files, including calibration and characterization data. The combination of Dolt for calibration data and MongoDB for characterization data emerged as the best-fit solution in our design exploration. Our approach ensures a customized and effective solution tailored to the team's specific challenges. QubiCSV is designed to address both data management and visualization aspects of quantum research, mainly focusing on calibration and characterization data.

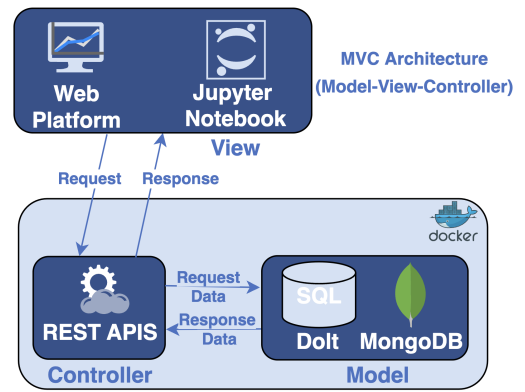

**Figure 3.** QubiCSV's system architecture.

The system is structured into three major components: (1) **Application Programming Interface (API)**: It acts as the communication bridge between the database and the user interfaces, overseeing the transfer and retrieval of data to ensure smooth interaction among the system's different components[33]. (2) **Web Platform**: A user-friendly web application serves as the primary interface for users, facilitating various tasks and actions such as data

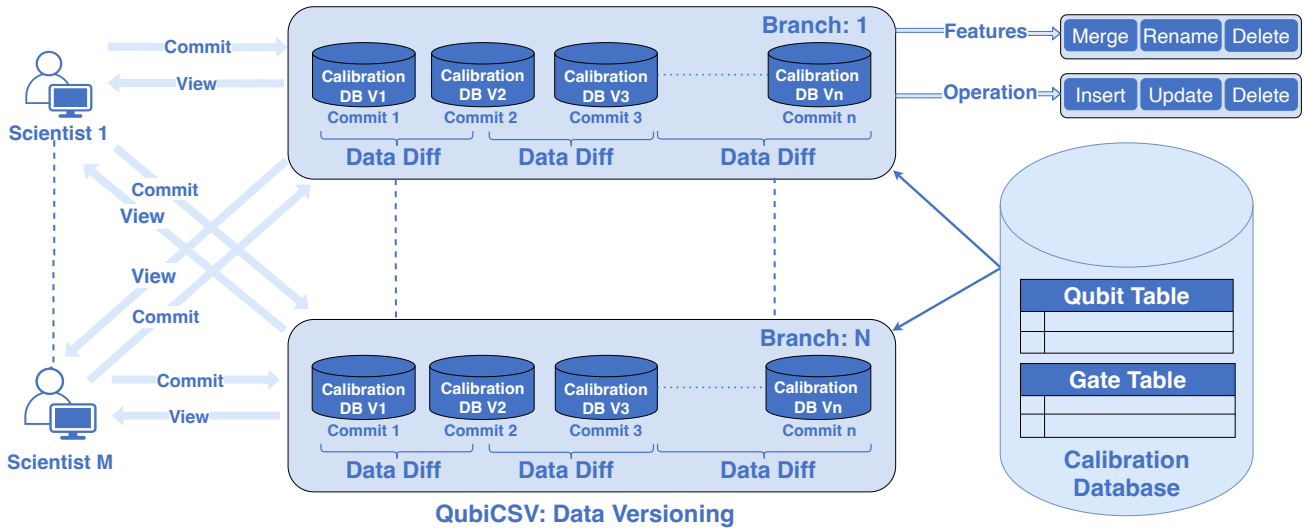

Figure 4. QubiCSV's data storage system with versioning capacity: This component of the system utilizes dolt, a data versioning database, enabling users to create and manage their own versions of databases in branches, similar to how git functions for source code. users can effortlessly create, access, and switch between different branches

management and visualization. (3) **Python Library**: Recognizing the team's reliance on Jupyter Notebook, we incorporated a Python library interface for seamless data storage and retrieval directly within their existing Jupyter Notebook workflows, as illustrated in Figure 3.

### 2.2 Calibration Data Management

#### 2.2.1 Calibration Data Versioning: Motivations and Approaches

Our platform leverages the data versioning database for calibration files, which is designed to address a few key needs:

(1) **Robust Tracking and Versioning:** Given the regular updates and numerous versions of calibration files used in various experiments, there is a crucial need for robust tracking and versioning mechanisms. The decentralized nature of the system also complicates collaboration and sharing, as it hinders the ability to track changes or access different versions in a straightforward manner.

(2) **Centralized and Collaborative Access:** Maintaining and updating calibration files is a daunting task, especially as the frequency of calibration increases. Scientists, each focusing on different aspects of the quantum system, generate their own calibration files which need to be accurately tracked and updated. The need for collaboration among team members further adds to the complexity. Ensuring access to the latest versions of these extensive files is a logistical challenge, impeding efficient collaboration.

QubiCSV provides a centralized storage solution with versioning capabilities, allowing each team member to not only track their work but also access and contribute to others' work seamlessly, as shown in Figure 4. As an example, we observe a scenario where multiple scientists (Scientist 1 to Scientist M) interact with the database. Each scientist has the capability to create multiple branches and access branches created by others. This flexibility in accessing and contributing to different branches fosters collaborative research and data sharing. Within each branch, scientists can maintain their unique versions of the database. They have the freedom to perform various operations like inserting new data, updating existing data, and deleting data. Moreover, they can merge two different branches, enabling the combination of data sets and collaborative developments. Branches can also be renamed or deleted as per the evolving needs of the research, ensuring the database remains organized and relevant. These data versioning capabilities are important for the management of calibration data in quantum computing research. They allow scientists to track changes, revert to previous data states if necessary, and collaborate effectively with peers. If a conflict occurs, users are notified and can choose between retaining the modified or unmodified version, similar to Git, Dolt requires users to resolve cell-level conflicts when merging data, offering strategies like 'ours' or 'theirs,' or allowing manual resolution. The system's design, emphasizes not just the technical capability of data versioning but also its practical application in a research environment, making it an invaluable tool for scientists working with quantum systems.

#### 2.2.2 Calibration Data Schema Design

Calibration data of qubit configuration are managed in a JavaScript Object Notation (JSON) format, which can be either in a JSON file or as an individual object in a Jupyter Notebook. This quantum computing calibration dataset details the parameters governing qubits and gates within the quantum processor. Key attributes include the qubit drive frequency (*freq*) and qubit readout frequency (*readfreq*). Additionally, the file outlines gate configurations. For instance, the X90 Gate for Q0, specifying frequency (*freq*), phase (*phase*), destination (*dest*), time width (*twidth*), start time (*t0*), amplitude (*amp*), and an envelope function (*env*) like "cos_edge_square" with a 25% ramp fraction for rising and falling edges. This comprehensive configuration extends to various qubits, providing detailed settings for each in terms of their drive and read frequencies, gate operations, and associated envelope functions.

1. Data Sample: Calibration data

```
"Qubits": {
  "Q0": {"freq": 4100733234.438625,
      "readfreq": 6554300000.0,
      "freq_ef": 4182558902.85729}},
"Gates": {
  "Q0X90": [{"freq": "Q0.freq",
      "phase": 0.0,
      "dest": "Q0.qdrv",
      "twidth": 3.2e-08,
      "t0": 0.0,
      "amp": 0.50617256269105,
      "env": [{"env_func": "cos_edge_square", "paradict": {"ramp_fraction":0.25}}]}]}}
```

The database schema of our platform is intricately designed to accommodate the specific needs of calibration data. Utilizing Dolt, a data versioning system built upon a traditional SQL database structure, we have established a schema that efficiently organizes calibration details for every chip. In this schema, the chip acts as a foreign key in both the gate and qubit tables. The detailed structure of our database schema is illustrated in Figure 5.

### 2.2.3 Implementation of Calibration Data Management

A dashboard resembling a traditional Git interface facilitates data management, as illustrated in Figure 6. These screenshots provide a depiction of how users interact with the system. The interface includes a home page that serves as the entry point to the platform. From there, users can navigate to a dashboard specifically designed for managing branches, exploring individual branches or viewing details of specific commits. Additionally, the platform offers a feature to compare two commits, allowing users to easily identify differences and track changes over time. This visual overview underscores the platform's user-friendly design and functionality in managing and visualizing calibration data in quantum computing research. The key functionalities are detailed in Table 1, which outlines the comprehensive features and capabilities of the calibration data management system within the QubiCSV platform. The table also includes code snippets for each functionality, providing users with practical examples of how to utilize these features within the Python library.

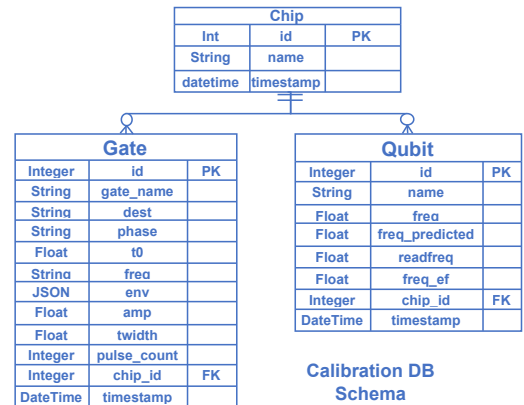

Figure 5. The schema for calibration data consists of three primary tables: chip, qubit, and gate. In this schema, each 'chip' is designated as a primary key (PK) and serves as a foreign key (FK) in both the qubit and gate tables. This setup allows one chip to be associated with multiple qubits and gates.

## 2.3 Characterization Data Management

Characterization data is a critical output generated after the calibrated QubiC system measures qubits from quantum computers. This data file contains various properties for each qubit, offering valuable insights into their performance. The key properties include *prep0read1, prep1read0, rb1qinfidelity, separation, t1, t2ramsey, and t2spinecho*. Format of characterization data file typically follows the naming convention *Chip_name.data.json*. In our platform, the chip name is extracted directly from the filename, streamlining the process of data identification and retrieval.

1. Readout Fidelity (*prep0read1*, *prep1read0*): The *prep0read1* and *prep1read0* metrics offer insights into the fidelity of state preparation and measurement, which are crucial for ensuring the reliability of quantum operations.

2. Randomized Benchmarking Infidelity (*rb1qinfidelity*): The *rb1qinfidelity* metric is a key indicator of the infidelity of single-qubit gates. It plays a crucial role in assessing the quality and precision of quantum operations, providing insights into how accurately these gates can manipulate the state of a qubit without introducing significant errors.

3. Qubit readout *Separation*: *Separation* is the distance between qubit blobs, which shows how clearly different qubit states can be identified from each other.

| Features | Function Description | Code Example |
|---|---|---|
| Create Branch | *Users can create a new branch by specifying the branch name, owner's name and email, and a description. After creating a branch, they can select the appropriate chip and upload the calibration file. Physicists have the flexibility to create multiple branches and access any existing branches* | calibration.createbranch (commit_data, branch_name) |
| Merge Branch | *Physicists can execute merges by specifying details such as from_branch, to_branch, owner, and a message. Upon initiating the merge, all the data from the from_branch is seamlessly integrated into the to_branch, resulting in the creation of a new merge commit* | calibration.mergebranch (from_branch, to_branch, author_name) |
| Rename Branch | *This feature enables the modification of branch names as needed, accessible both through the UI and the Python library. This flexibility allows physicists to change the branch name to suit their current experiment or specific requirements* | calibration.renamebranch (old_branch_name, new_branch_name, author_name) |
| Copy Branch | *Mirroring the 'clone' function found in Git, this feature enables users to replicate an entire branch into a new branch. This capability is particularly useful for preserving the original data state while experimenting with variations in calibration.* | calibration.copybranch(branch_name) |
| Delete Branch | *This feature allows for the removal of unwanted data branches. This action requires specifying the branch name for confirmation, ensuring that branches are not deleted unintentionally or without proper authorization* | calibration.createbranch (branch_name, author_name) |
| History of Repository | *This function provides a comprehensive log of all activities at the database level. It meticulously tracks key actions such as the creation, deletion, or renaming of branches, as well as recent commits* | calibration.history( ) |
| Commit Data | *In our system, users upload a calibration file for a chosen chip, followed by a commit operation where they enter author details and a commit message. Successful uploads generate a SHA-256 hash as a unique commit identifier.* | calibration.insertbyfile(file_path, commit_data, branch_name, chip_id) |
| View Calibrated Data | *After committing a calibration file, users can instantly view it on the interface and access a table of their data. In Jupyter notebooks, a JSON file named after the commit hash (e.g., Commit_Hash.json) is generated, containing the full commit data.* | calibration.getcommitdetails (commit_hash, branch_name) |
| Data Diff | *This tool allows physicists to compare commits within a branch, visualizing calibration data differences. By using the Data Diff function, they can closely inspect changes across commits, aiding in identifying the best calibration combination* | calibration.getcommitdiff (commit_hash, branch_name) |

Table 1: Comprehensive overview of QubiCSV data versioning features for calibration data management. It includes feature descriptions and Python code examples for easy implementation in Jupyter Notebooks. This design allows seamless database interactions from notebooks, mirrored on the platform's dashboard.

4. Coherence Times (*t1*, *t2ramsey*, *t2spinecho*): The coherence time refers to the duration during which a qubit can retain its quantum state, essentially representing the lifespan of a qubit. The *t1* measures the time it takes for a qubit to relax to its ground state, while *t2ramsey* and *t2spinecho* gauge the qubit's dephasing time, reflecting how long it maintains its quantum state coherently.

For long-term monitoring data, specifically the characterization files crucial for understanding experiment outcomes, we opted for a NoSQL MongoDB [34] database. This decision was guided by the nature of the experimental result files, which are stored in JSON format. MongoDB, being a NoSQL database, naturally supports JSON data, making it an obvious choice for our requirements. Its flexible schema and powerful querying capabilities greatly facilitate the utilization of these JSON files for visualization purposes. Furthermore, MongoDB's scalability and performance efficiency make it an ideal fit for handling the extensive datasets typical in quantum experiments. This setup in MongoDB enables the efficient tracking of all experimental results for each qubit, thereby maintaining a comprehensive record of research progress. MongoDB can also handle high volumes and can scale both vertically or horizontally to accommodate large data loads, due to its scale-out architecture[35].

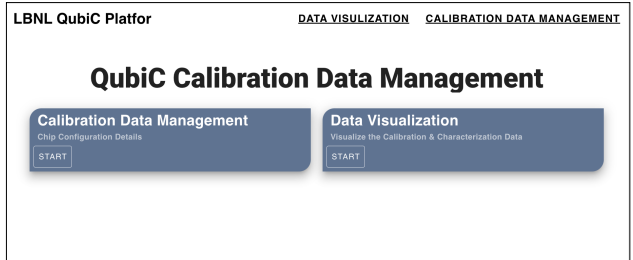
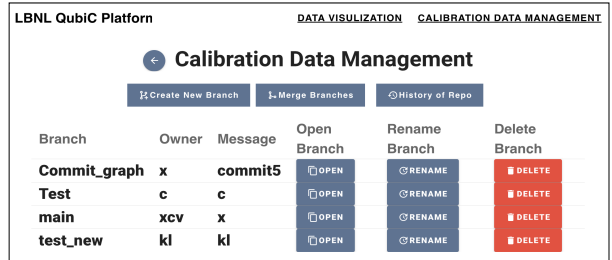
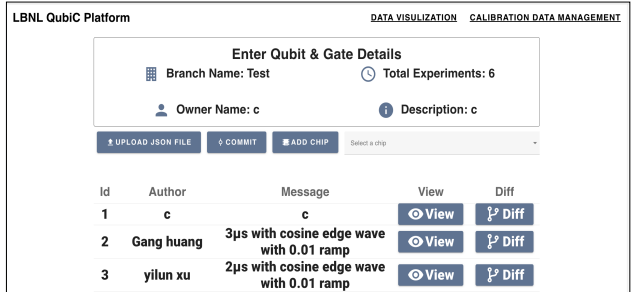
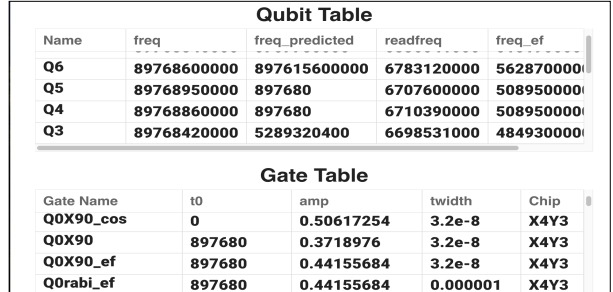
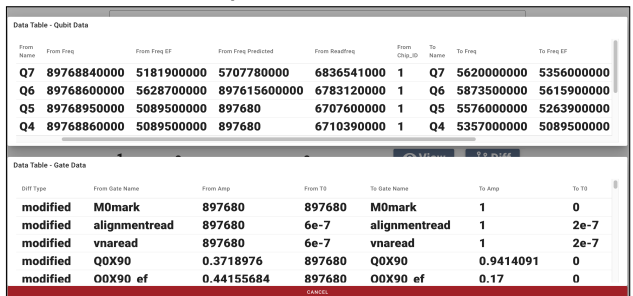
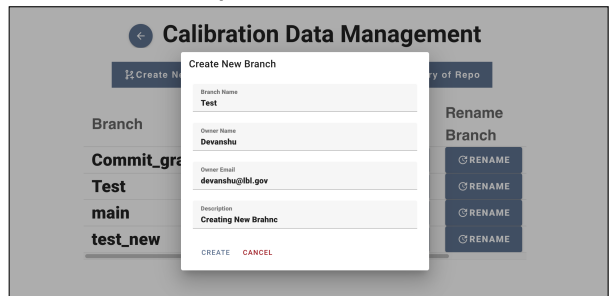

**Figure 6.** QubiCSV's dashboard screenshots. This image presents a collection of screenshots showcasing the user interface of the QubiCSV platform, providing a visual overview of the system's features and user interactions.

2. Data Sample: Characterization Data

```
{ "Q0": { "prep0read1": {
        "enddatetime": "20220526_182729_656142",
        "mean": 0.00290175,
        "startdatetime": "20220526_180730_062549",
        "std": 1.8719794650678421},
    "t2spinecho": {
        "enddatetime": "20220526_182729_656142",
        "mean": 8.3675e-05,
        "startdatetime": "20220526_180730_062549",
        "std": 6.59268344454669e-06}}}
```

## 2.4 QubiCSV Data Visualization

Visualization plays a crucial role in improving the comprehension and monitoring of quantum experiments. It allows researchers to observe and analyze complex Quantum calibration and characterization data in an intuitive manner. Through visualization, patterns and insights that might otherwise remain obscured in raw data can be brought to light, facilitating a deeper understanding of quantum phenomena. In the context of QubiC, effective visualization tools can transform how scientists interact with and interpret Quantum calibration and characterization data, leading to more efficient and insightful experiments. From our thorough analysis, Vuetify[36] web framework is selected for its sleek design capabilities and ease of integration. Vuetify's extensive component library allowed us to create a user-friendly interface, which simplifies the complex process of analyzing and interpreting calibration data, thereby enhancing the understanding of the quantum system's performance.

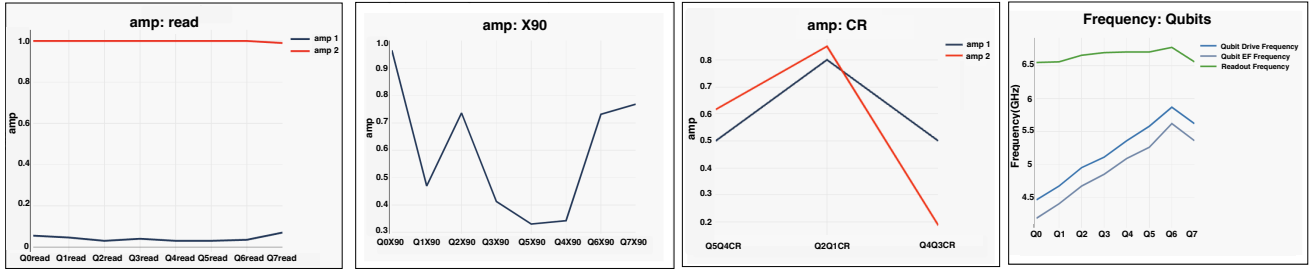

Figure 7. Visualization of gate groups & qubits for a specific commit in QubiCSV. Detailed readings of amplitude values for all read gates, X90 gates, and CR(cross resonance) gates are showcased for a selected commit. The chart also displays the qubit drive frequency, qubit e-f transition frequency, and readout frequency for all qubits, offering insights into their operational characteristics for the selected commit.

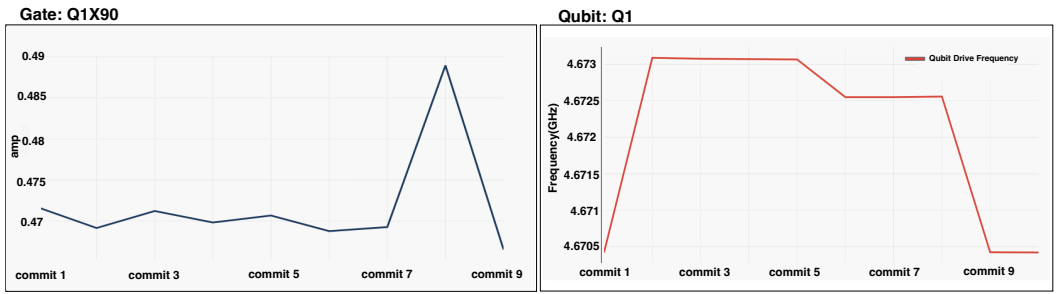

Figure 8. Detailed frequency and amplitude visualization in QubiCSV. This figure illustrates the X90 gate amplitude, and qubit drive frequency values for Q1 across all commits, offering a view of specific qubit and gate behavior over time.

For visualizing experiment results, the key requirement was a tool capable of rendering complex data plots dynamically. Plotly.js[37] emerged as the ideal choice for this purpose, due to its advanced graphical options and interactivity. It enables users to delve deeply into the experiment outcomes, facilitating a detailed and nuanced exploration of the data. To support this front-end capability, we needed a robust back-end solution. Here, Flask[38] was chosen for its simplicity and efficiency as a Python web framework. Flask's ability to handle data processing and API management made it the perfect match for our system's back-end requirements.

These decisions in our design process were driven by the need to balance functionality with user accessibility. Vuetify's user-centric design approach with automatic tree shaking, an easy-to-learn API, server-side rendering (SSR), progressive web app (PWA) support, and mobile app support, among other features. It also offers internationalization support, right-to-left (RTL) text support, and a blazing-fast framework experience. These features make Vuetify a versatile and efficient choice for building modern, high-performance web applications and Plotly.js's[37] advanced plotting capabilities, it is a declarative charting library that offers over 40 chart types, including sophisticated options like 3D charts, statistical graphs, and SVG maps. This flexibility makes it the best choice for creating interactive, high-quality visualizations, combined with Flask's back-end proficiency[39], culminating in a broad visualization system.

The database structure we adopt is designed to efficiently organize and store characterization data for each qubit. The structure is as follows: each entry is identified by an id and is associated with a specific qubit. The ExperimentData field is an array of objects, each representing a set of characterization data linked to a particular chips. When a user uploads a new experiment.json file, the system is designed to add the new characterization data to the respective qubits and append it to the ExperimentData object array. This approach ensures that all characterization data is systematically recorded and easily retrievable for each qubit, enabling comprehensive tracking and analysis over time.

### *2.4.1 Visualization of Calibration Data*

To visualize the calibration data, users first select the desired branch from the database. This step is crucial as our system incorporates data versioning, allowing for a detailed historical perspective of the data. After selecting a branch, users are presented with a list of all available chips, along with their respective properties[40]. From this point, users can choose which aspects of the calibration data they wish to explore visually. Our visualization feature offers two primary types of charts, each providing unique insights into the calibration data:

■ *Charts By Commit (Branch and Chip Specific):* This type of chart allows users to visualize all qubits and gates characteristics for individual commits within a selected branch and chip. It provides a snapshot of specific calibration states over time, enabling users to track changes and identify trends or anomalies in the calibration process. We visualize key qubit characteristics such as readout frequency, qubit drive frequency, and qubit e-f transition frequency. Moreover, we showcase essential gate characteristics including phase, frequency, amplitude, and time width. An example of these charts is shown in Figure 7. For instance, the graph clearly displays the amplitude values of all X90 gates for a particular commit, offering insight into the intensity of the signal used in gate operations. Similarly, the chart includes values for readout and CR gates. Additionally, the visualization extends to qubit frequency values, where the chart showcases the frequencies of all qubits for the same commit. This comprehensive display of both gate and qubit values at a specific commit point is important in understanding the calibration process's intricacies and effectiveness:

Considering the wide variety of gates, we've grouped them into categories based on similar characteristics. This methodology not only enhances the organization of the data but also significantly improves the clarity and effectiveness of our visual analysis. For example, we categorize all readout gates into the *ReadGroup*, all 90-degree rotation (along the X-axis) gates into the *X90Group*, and all two-qubit CR gates into the *CRGroup*.

■ *Charts By Properties (Commit Specific):* For a given property, these charts display all the commit values for individual gates and qubits. This approach is especially valuable for examining the evolution of specific properties across different commits, providing a deeper insight into the dynamics of the calibration data. Figure 8 shows an example of these charts, which provides a comprehensive view of how specific properties change over time for each commit.

- *Qubits*: We have detailed property-specific visualizations for qubits. This visualization approach provides individual property charts for each qubit, showcasing how properties like qubit drive, e-f transition, and readout frequency have evolved over different commits.
- *Gates*: Similarly, for gates, these graphs offer a comprehensive view of the evolution of gate properties such as phase, frequency, amplitude, and time width over time. Users can analyze these property-specific charts for each gate, enabling them to closely monitor the behavior and performance of the gates across various commits. This level of detailed visualization aids in identifying patterns, trends, and potential areas for optimization in the gate operations.

This approach demonstrates how we plot graphs for each gate, showcasing the dynamic nature of our system. It is designed to seamlessly accommodate new gates and qubits as they are added to the calibration file. With data versioning and visualization, QuiCSV provides the most effective monitoring and analysis capabilities to scientists engaged in quantum research work. This feature enhances the flexibility and adaptability of our platform, ensuring it remains a valuable tool in the ever-evolving field of quantum computing.

### *2.4.2 Visualization of Characterization Data*

Characterization Data, generated after calibrated QubiC interacts with qubits, plays a pivotal role in comprehending the effectiveness of quantum experiments. It offers insights into how each qubit responds to calibration and reveals patterns in experimental behaviors. Identifying these patterns is crucial for determining the optimal calibration combinations that yield the best results. Our platform provides two distinct approaches to visualize this characterization data, catering to different analytical needs. To begin visualization, users must first select the specific chip whose characterization data they wish to analyze. This initial step ensures that the subsequent data visualizations are tailored to the selected chip (Figure 9).

- *By Qubits*: This method concentrates on a particular qubit, enabling users to monitor and analyze all properties associated with that qubit across various experiments. Researchers can scrutinize changes and trends in properties such as prep0read1, rb1qinfidelity, t1, t2ramsey, and others, across different experiments. Furthermore, researchers can pinpoint specific conditions under which the qubit operates optimally or displays anomalous behavior.
- *By Properties*: Alternatively, users can choose to focus on a specific property and observe how all qubits respond to this property across multiple experiments. This approach is particularly useful for analyzing how a specific property, like coherence time or readout errors, varies across different qubits and for identifying patterns or anomalies that are consistent or variable across the qubit array.

Both these visualization methods are designed to provide in-depth insights into the experimental data, assisting researchers in making informed decisions about future experiments and calibrations.

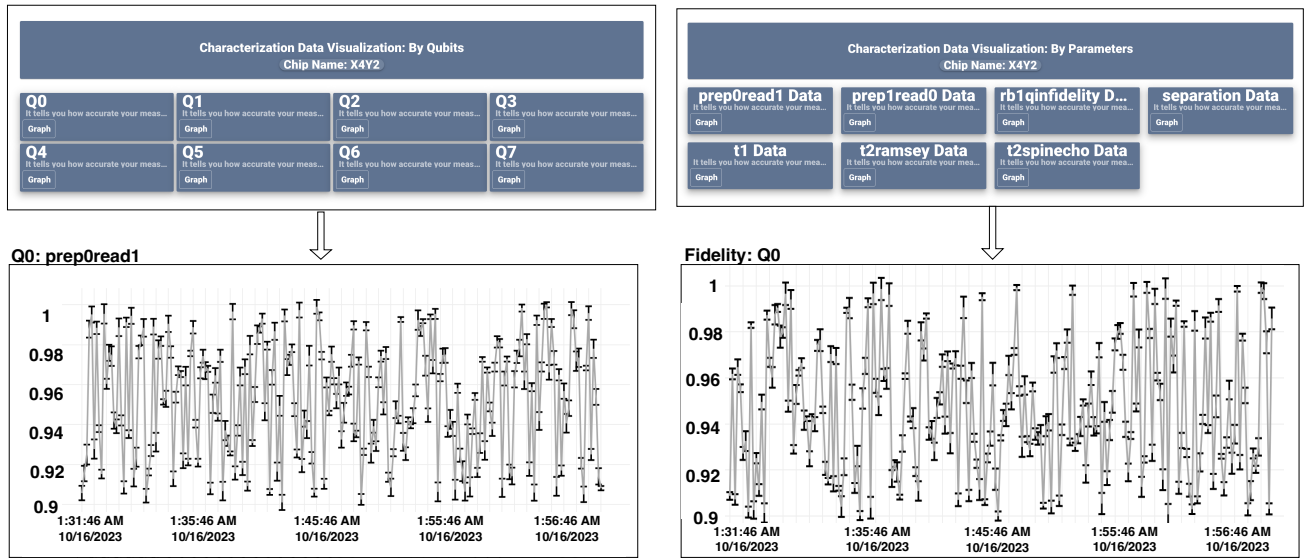

**Figure 9.** Examples visualization of data by qubits and properties.

*2.4.3 User Accessibility Interaction Interface*
In the context of QubiCSV, 'User Accessibility Methods' refers to the various ways in which users can interact with and access the platform. The users interact with the platform to manage and visualize quantum computing data, specifically calibration and characterization data. The frequency of interaction varies, with some users accessing the system daily for active experiments while others might use it less frequently for data review or research purposes. The design of the QubiCSV user interface was a complex task, particularly given the diverse needs and preferences within the scientific community. Initially, our primary focus was developing a Python library, considering scientists' widespread use of Jupyter Notebook for managing calibration files.

As we aimed to make QubiCSV an open-source platform accessible to a broader range of users including other research facilities, educational institutions, and individual researchers, we recognized the need for a more inclusive approach. A Python library, while efficient, might not cater to all potential users, especially those who are not as familiar with coding or prefer a more interactive interface. This led us to consider the advantages of a web-based platform, which could offer a more user-friendly and visually intuitive experience. A web-based interface would not only be beneficial for new or less technically inclined staff but also for the wider research community who might prefer a more graphical interface for data visualization and management.

Consequently, we opted for a dual-interface approach. We introduced a web-based platform to enhance the user experience with advanced visualization capabilities and a more intuitive interface. Simultaneously, we developed a Python library that seamlessly integrates with Jupyter Notebook. This library facilitates the same API calls for storing and retrieving data from the database, ensuring that users comfortable with Jupyter Notebook can continue to work within their preferred environment. This dual-interface approach ensures that our platform accommodates the varying needs and preferences of the scientific community. This design decision, therefore, not only caters to the immediate needs of LBNL scientists but also positions QubiCSV as a versatile tool for the broader quantum computing research community.

## 2.5 Contribution & Performance
Our platform offers sophisticated storage and visualization capabilities for each calibrated gate and qubit. For every data insertion, we provide detailed and interactive charts. These plots feature built-in screenshot capabilities, zoom-in-out, panning, and auto-scaling, catering to the diverse needs of data analysis.

The platform boasts an API response time of less than 500 ms. The adoption of the MVC architecture offers flexibility, particularly in modifying the user interface. This architectural choice means that if someone wishes to use our APIs to create their own visual board with modifications, they can easily do so, allowing for customization and adaptability to individual research needs. Regarding the performance of database versioning, we base our assessment on comparisons with MySQL using standard sysbench[41] metrics. Dolt, being MySQL compatible, provides a relevant benchmark for performance evaluation. Currently, Dolt's performance is approximately 1.9X slower than MySQL: 1.3X slower for write operations and 2.3X slower for read operations, as per the standard suite of sysbench tests. Despite being slower than MySQL, with most MySQL queries returning in the 0-10ms range, Dolt's performance remains within a practical range for user applications, especially considering its versioning capabilities.

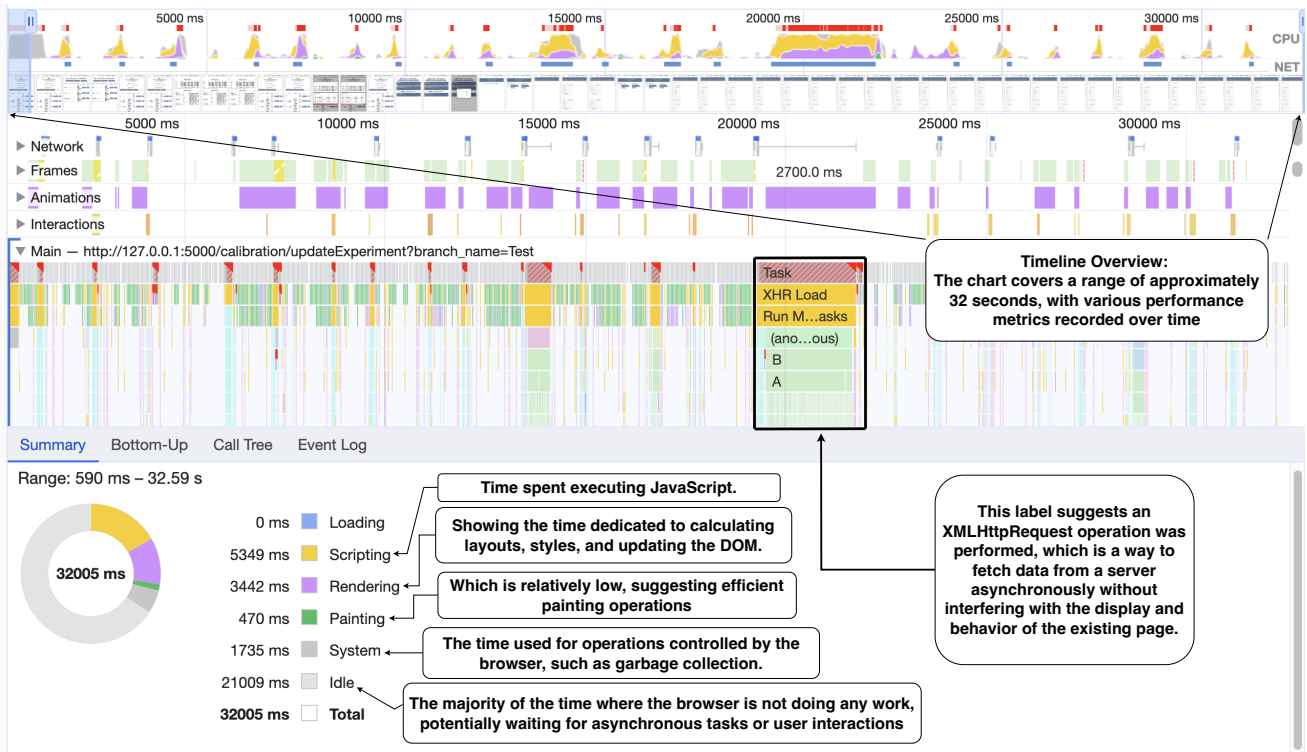

Figure 10. Over a 30-second period under a 4x CPU slowdown, the platform demonstrates swift handling of complex visualizations with Plotly, maintaining fast scripting (5129 ms) and rendering (3375 ms) amidst substantial idle time, indicating a responsive system optimized for heavy data operations, it demonstrates the web platform's efficiency in managing complex visualizations and API data fetching, maintaining responsiveness and good performance metrics obtained from Google Chrome.

This Chrome Developer Tools Performance tab screenshot captures a detailed performance profile of a web platform as shown in Figure 10, especially under a 4x CPU slowdown simulation, which is often done to mimic less powerful devices and ensure that the web application performs well even on lower-end hardware. The performance analysis of our web application over a 30-second period reveals a timeline of color-coded activities, each representing different browser tasks—scripting in yellow, rendering in purple, painting in green, and system operations in gray—with red triangles highlighting long tasks that may impede responsiveness. A detailed flame chart illustrates the JavaScript call stack, pinpointing performance bottlenecks, while network requests indicate the timing of data transmission crucial for API performance assessment. The frame rate and rendering times show the efficiency of style recalculations and screen drawings. The summary pie chart at the bottom details the time allocation across operations, with the majority being idle time (21,090 ms), suggesting the system awaits completion of tasks or user input. Notably, the scripting (5,349 ms) and rendering (3,442 ms) times remain low compared to the idle time, which indicates that the web application is efficiently handling the heavy visualization and chart plotting using Plotly, even with the increased load from a slower CPU.

## 3 Discussion

One of the standout features of our platform is its function as a comprehensive data management system tailored for QubiC, one of the few open-source qubit control systems worldwide. QubiC, being an experimental tool, produces a vast amount of data. Before our platform, there was a lack of a suitable system to store and efficiently utilize this valuable data, which contains invaluable insights crucial for quantum research. QubiCSV addresses the intricate challenges associated with managing calibration and characterization data, two critical components in the realm of quantum computing. Our commitment to open-source development and ongoing user feedback reinforces QubiCSV's position as an invaluable tool in quantum research, meeting the dynamic demands of QubiC team and enhancing data analysis and visualization capabilities.

### 3.1 Integration of QubiCSV with QubiC

The open-source nature of the QubiC system, combined with our initiative to make QubiCSV open-source as well, provides a novel and effective solution for the community. This integration allows users not only to store large-scale experimental data efficiently but also to engage in detailed data analysis and visualization. This comprehensive approach enhances the understanding of quantum experiments and accelerates the research process. Researchers using QubiC can now easily access and comprehend data and experimental progress through QubiCSV, avoiding the time-consuming task of building their own tools or navigating through raw files.

### 3.2 Upgrading QubiCSV to work with other systems

QubiCSV's open-source nature makes it a prime candidate for adaptation by others. The platform's architecture, which includes the Dolt database for calibration data and MongoDB for characterization data, is accessible via our gitlab repository[42]. A notable aspect of the setup process is the inclusion of a Docker file, complete with comprehensive setup guidelines. This Docker file simplifies the installation process, enabling users to quickly host and run the entire project with just a few commands.

This setup, combined with the MVC architecture for API and user interface (UI) design, is clearly documented, allowing easy replication or modification. The tutorial for users to start the QubiCSV system has been made available online[43].

- *Database Schemas*: The schemas for both Dolt and MongoDB are provided in the code scripts, allowing easy customization based on different calibration and characterization data needs.
- *API Structure*: Adjustments to the REST API may be needed to align with different data structures or user requirements.
- *User Interface Customization*: Depending on the specific application, the user interface might require modifications, which can be done within the MVC framework.

Our comprehensive documentation serves as a valuable resource for anyone looking to adapt QubiCSV. It provides detailed guidance on setup, architecture, and customization.

### 3.3 Limitations & Future work

While QubiCSV offers valuable functionalities, there are certain limitations that need to be addressed for its further enhancement. The current visualization capabilities require improvement to provide users with more intuitive and insightful data representation. Looking ahead, collaboration with other researchers is seen as a promising avenue for expanding QubiCSV's capabilities and ensuring its adaptability across diverse platforms. Moreover, there is a keen interest in integrating machine learning tools into the database for long-term characterization data analysis, with a focus on providing feedback and mitigating qubit drift caused by environmental factors. These advancements are expected to significantly contribute to the robustness and effectiveness of QubiCSV in supporting quantum computing research endeavors.

## 4 Methods

We studied the limitations of existing quantum control design, one of the most important research topics in the current state of quantum superconducting qubit research. We then come up with new design goals and technical challenges to bridge the gaps. We implemented novel data versioning and visualization techniques, allowing multiple researchers to interact and collaborate on the same qubit hardware.

### 4.1 Design Goals

We identified a few key requirements of QubiCSV design as follows:

- *Collaborative Platform*: There is a strong need for a platform that enables collaborative efforts, allowing physicists to work together and contribute to each other's work seamlessly from the user study that we conducted with LBNL team. This quantum research team is characterized by its rich diversity, encompassing interns, postdocs, control engineers, and staff physicists, each bringing unique perspectives and requirements to the table.
- *Tracking and Versioning*: Frequent updates and multiple versions of calibration for various experiments highlighted the necessity for robust tracking and versioning capabilities. For example, when team members, such as interns or postdocs, require a specific calibration file, they must request it directly from the physicists responsible for its creation. This process can be time-consuming and inefficient, as physicists typically maintain their calibration files locally without a centralized storage system. Consequently, there is no effective way to track these files' evolution or individual experiments' progress. Moreover, the lack of a centralized system makes sharing and collaborating on these files difficult, as there is no straightforward method to track changes or access different versions. This inefficiency in the current practice not only delays experimentation but also hampers the collaborative potential of the team.

- *Visualization of Calibration and Experiment Outcomes*: The team's objective is to achieve optimal qubit performance through the best possible calibration configurations for their experiments. To facilitate this, they required detailed visualization plots for both calibration data and the outcomes of experiments conducted with that calibrated data. By analyzing the visualized data, the team aims to identify the most effective calibration configurations, focusing on parameters such as the frequency and amplitude settings for gates and qubits. This process of visualization and analysis is crucial for fine-tuning the calibration settings, ultimately leading to enhanced performance and efficiency in quantum experiments.
- *Dual Interface Design*: We observed that quantum research team members primarily use Python Jupyter Notebook[44] for managing JSON calibration and characterization data, as well as for coding. This established practice highlighted the need for a dual interface design in our new system. While there was a clear preference for continuing with the familiar Jupyter Notebook interface, there was also a strong desire for a more intuitive, dashboard-like interface to enhance the overall user experience.

### 4.2 Technical Challenges

Developing QubiCSV platform faces many challenges.

- *Complex Data Management*: Quantum research groups, such as the QubiC team at LBNL, working on various system components, generate substantial calibration data that is both extensive and constantly evolving. The management and updating of these calibration files, integral to the functionality of quantum systems, pose a daunting task for researchers.
- *Collaboration Hurdles*: Sharing extensive calibration files and ensuring access to their latest versions adds complexity, hindering efficient collaboration among team members.
- *Lack of Effective Data Storage and Analysis*: Post experimentation, systems like QubiC produce valuable experiment result files (e.g., chip_name.data.json), which contain configurations obtained through multiple time-consuming iterations, but the absence of a dedicated storage and direct saving method from the hardware limits their utility.
- *Time-Intensive Calibration Setting*: There is limited data analysis or visualization in the existing quantum research interface. Without data analysis, much time is spent on setting the best values for each calibration property. Visualization tools can significantly streamline this process, helping to achieve optimal calibration with reduced error and decoherence.
- *Scalability and User-Friendly Design*: QubiCSV is designed to be scalable and loosely coupled. As the number of qubits and gates increases, our system can effortlessly adapt to these changes. We have also provided a well-documented user manual, accessible at QubiCSV Documentation[43], which guides users on how to set up and utilize the platform effectively.

### 4.3 Designing and Implementing Data Management and Versioning

Initially, we considered using a MySQL database with versioning implemented through timestamps and tags. However, the need for a more dynamic system capable of handling frequent updates and multiple versions of calibration data steered us away from this path. The SQL model's rigidity in versioning and the complexity involved in implementing custom versioning solutions were significant drawbacks. Our exploration then led us to the innovative data versioning database called "Dolt". Dolt, operating similarly to Git but for data, offers functionalities to fork, clone, branch, merge, push, and pull a SQL database as if it were a Git repository. This choice addressed our need for a flexible and user-friendly version control system, allowing users to manage calibration data with the same ease as source code. This capability of Dolt perfectly aligned with our requirements. Additionally, Dolt has demonstrated its performance[45] through standard tests like sysbench[46], where it competes closely with MySQL in terms of latency. Though Dolt is approximately 2X slower than MySQL, with 1.5X on writes and 2.5X on reads, it still offers competitive performance, especially for our requirement of versioning large calibration datasets. This benchmark, reflecting industry-standard online transaction processing (OLTP) oriented tests, justifies our choice of Dolt. It offers a balance of familiarity in operations akin to Git with a performance level that, while slightly slower, is adequately suited for the complex data management needs of quantum computing research.

### 4.4 Designing and Implementing Data Visualization

In addition to offering chart plotting for individual commits, our platform provides detailed visualizations for each property within a commit, such as frequency, phase, amplitude, and time width. This feature allows users to closely examine the changes and trends in specific properties over time. Additionally, we offer charts that compile data across all commits for individual properties, providing a broader view of data evolution. This dual approach to visualization—both at the individual commit level and across multiple commits—ensures that users gain a deeper and more nuanced understanding of the calibration data, facilitating more informed decision-making and analysis.

## 5 Data Availability

The main codebase of our project is hosted at our GitLab repository. This public repository includes the functionality for database connection, data versioning, and contains the frontend code: https://gitlab.com/DevanshuBrahmbhatt/

`qubic-data-storage`. For code access and deployment, we utilize Docker. The deployment scripts, designed for local setup, can be found at our deployment's repository. The platform operates locally on port 5000, and the deployment script will deploy the project in your local environment: https://gitlab.com/DevanshuBrahmbhatt/qubic-deployments. Additionally, for users who prefer to interact with our platform via Jupyter notebooks, we have a dedicated repository which can be found here: QubiCSV Jupyter notebook repository: https://gitlab.com/DevanshuBrahmbhatt/qubic-jupyter. To assist new users with getting started, we have a comprehensive user manual available at this GitBook URL. This guide details the steps for accessing and utilizing the QubiCSV platform effectively: https://devanshus-organization.gitbook.io/qubic-docs.

Please note that our platform is specifically tailored for the storage and visualization of calibration and characterization data. We do not use external datasets; instead, we focus on visualizing the data that we store. The links to the calibration and characterization files which are stored in the database that we are sharing. We are not using any existing dataset, but we are sharing the original data (calibration.json and characterization.json) that we are storing in our database. The calibration.json and characterization.json are included in the related files.

The datasets used and/or analyzed during the current study are available from the corresponding author on reasonable request. All data generated or analyzed during this study are included in this published article and its supplementary information files. The datasets generated and/or analyzed during the current study are not publicly available because these calibration and characterization data changes by qubit chips but are available from the corresponding author on reasonable request. Sample data are made publically available: Calibration file[47] and Characterization file [48]. There is no data from a third party. Our study does not involve the use of hospital or health-related data. Therefore, the requirements mentioned regarding institutional and/or licensing committee approval, as well as informed consent from subjects or their legal guardians, are not applicable to our study.

## 6 Conclusion

We have thoroughly examined the constraints of existing Quantum calibration and characterization data management practices and successfully developed and implemented a comprehensive data storage and visualization system known as QubiCSV. Our platform not only facilitates data storage but also empowers users to generate various plots and visualizations, aiding scientists in deriving meaningful insights from their experiments. A key feature of QubiCSV is its data versioning capability, which enhances collaborative research by enabling multiple versions of calibration and characterization data to be managed effectively using the Dolt and MongoDB databases for qubit control. Furthermore, QubiCSV has been deployed on a server and seamlessly integrated into QubiC, demonstrating an exemplary showcase of a more accessible and collaborative research platform in the field of quantum computing.

## Acknowledgements

This work was supported by the U.S. Department of Energy, Office of Science, Office of High Energy Physics, and the National Quantum Information Science Research Centers Quantum Systems Accelerator under Contract No. DE-AC02-05CH11231.